\documentclass[10pt,conference]{IEEEtran} 

\usepackage{amsmath,amsfonts}
\usepackage{algorithmic}
\usepackage{graphicx}
\usepackage{textcomp}
\usepackage{xcolor}
\usepackage{url}
\usepackage{microtype}
    
\usepackage{xcolor}
\usepackage{xspace}

\newcommand{\eg}{\hbox{\emph{e.g.,}}\xspace}
\newcommand{\ie}{\hbox{\emph{i.e.}}\xspace}

\PackageWarning{TODO:}{X}
\PackageWarning{TODO:}{X\%}

\usepackage{tcolorbox}

\usepackage{tcolorbox}

\usepackage{ragged2e}
\usepackage{enumitem}
\usepackage{amsmath}
\usepackage{balance}
\usepackage{booktabs}
\usepackage{pifont}
\usepackage[ruled,vlined]{algorithm2e}
\usepackage{subcaption}

\usepackage{tcolorbox}

\usepackage{xspace}

\usepackage{xcolor}

\newcommand{\SEVs}{SEVs\xspace}
\newcommand{\SEV}{SEV\xspace}
\newcommand{\Meta}{Meta\xspace}

\newcommand{\StarBert}{\texttt{StarBERT}\xspace}

\title{Moving Faster and Reducing Risk: \\ Using LLMs in Release Deployment}

\author{
    \IEEEauthorblockN{Rui Abreu, Vijayaraghavan Murali, Peter C Rigby,\textsuperscript{*} Chandra Maddila, Weiyan Sun,}
    \IEEEauthorblockN{Jun Ge, Kaavya Chinniah, Audris Mockus, Megh Mehta, Nachiappan Nagappan}
    \IEEEauthorblockA{Meta Platforms, Inc., Menlo Park, USA}    
    \url{ruiabreu@meta.com}, \url{vijaymurali@meta.com}, \url{pcr@meta.com} \url{cmaddila@meta.com}, \url{wysun@meta.com}, \\ \url{jakege@meta.com}, \url{kanmanic@meta.com}, \url{audris@meta.com}, \url{meghmehta@meta.com}
}

\begin{document}

\maketitle

\begin{abstract}
Release engineering has traditionally focused on continuously delivering features and bug fixes to users, but 
at a certain scale, it becomes impossible for a release engineering team to determine what should be released. 
At \Meta's scale, the responsibility appropriately and necessarily falls back on the engineer writing and 
reviewing the code. To address this challenge, we developed models of diff risk scores (DRS) to determine how 
likely a diff is to cause a \SEV, \ie, a severe fault that impacts end-users. Assuming that \SEVs are only caused 
by diffs, a naive model could randomly gate X\% of diffs from landing, which would automatically catch X\% of 
\SEVs on average. However, we aimed to build a model that can capture Y\% of SEVs by gating X\% of diffs, 
where $Y >> X$. By training the model on historical data on diffs that have caused \SEVs in the past, we can 
predict the riskiness of an outgoing diff to cause a \SEV. Diffs that are beyond a particular threshold of risk 
can then be gated. We have four types of gating: no gating (green), weekend gating (weekend), medium impact on 
end-users (yellow), and high impact on end-users (red). The input parameter for our models is the level of gating, 
and the outcome measure is the number of captured \SEVs, \ie, the number of gated diffs that would have led to 
a \SEV. Our research approaches include a logistic regression model, a BERT-based model, and generative LLMs. 
Our baseline regression model captures 18.7\%, 27.9\%, and 84.6\% of \SEVs while respectively gating the top 
5\% (weekend), 10\% (yellow), and 50\% (red) of risky diffs. The BERT-based model, \StarBert, only captures 
0.61$\times$, 0.85$\times$, and 0.81$\times$ as many \SEVs as the logistic regression for the weekend, yellow, 
and red gating zones, respectively. The generative LLMs, iCodeLlama-34B and iDiffLlama-13B, when risk-aligned, 
capture more \SEVs than the logistic regression model in production: 1.40$\times$, 1.52$\times$, 1.05$\times$, 
respectively. 
\end{abstract}

{\let\thefootnote\relax\footnote{{\textsuperscript{*} Rigby is a professor at Concordia University in Montreal, QC, Canada.}}}
\section{Introduction}
Release engineering has focused on continuously delivering features and bug fixes to users. However, at a certain
scale~\cite{Rahman2016MSR,rossi2017rapid}, it is impossible for a release engineering team to determine what should 
be released. At our scale, the responsibility appropriately and necessarily falls back on the the engineer writing 
and reviewing the code. In this work, we develop models of diff risk scores (DRS), to determine how likely the diff, 
also known as a pull-request, is to cause a \SEV, \ie a severe fault that impacts end users. 

Assuming that \SEVs are only caused by diffs, a naive model could randomly gate $X\%$ of diffs from landing, \ie sent to the CI system for release, which would automatically catch $X\%$ of \SEVs on average. The question then is can we do better, \ie, can we build a model that can capture $Y\%$ of \SEVs by gating $X\%$ of diffs, where $Y >> X$. This is the value-add that machine learning (ML) models can bring to the table. By training the model on historical data on diffs that have caused \SEVs in the past, we can predict the riskiness of an outgoing diff to cause a SEV. Diffs that are beyond a particular threshold of risk can then be gated. Effectively, the model gives release engineers a knob that can be tuned to control productivity-risk trade-off.

We are also able to tune such model to be more or less conservative depending on the availability of engineers to deal with \SEVs and the potential impacting end users. We have four types of gating: no gating (green), weekend gating (weekend), medium impact on end users (yellow), high impact on end users (red). 
For example, on the weekend there are fewer engineers available to fix \SEVs leading to increased work for the on-call engineers, so diffs are gated to ensure that the author of the diff is confident that the change has been appropriately reviewed. 
During period of high use of our system, \eg around Black Friday, we want to ensure that we do not impact end-users as a result, diff landed during this period are gated at a very high level, \ie, 50\%, to ensure that engineers take an extra look at any diff that the model perceives to be risky.

We have one input parameter for our models and one outcome measure. The input is the level of gating and is done by picking the top X\% of risking diffs according to the model. Our {\it outcome measure} for a given model, m, and level of gating, g, is the number of captured \SEVs, \ie, the number of gated diffs that would have lead to a \SEV. We mine historical data to determine how well the model would have done compared to the one currently running in production. We group our models and model development around our research approaches. 
\subsection{Research Approaches}
\textbf{RA 1. Logistic Regression:}
How well does the current model capture SEVs?
%
Research into predicting bugs has shown a simple set of predictors to be highly effective~\cite{kdra20,dalal1988should,musa,GKMS99,MW00,nagappan2005use,KSAHMSU13,pachouly2022systematic,keshavarz2022apachejit,zhao2023systematic}. 
At the time of writing, the model used at \Meta has evolved to a logistic regression as it is know for its
robustness (not easy to get overfitted), efficiency to train and speed of prediction. As is common for just-in-time defect prediction~\cite{MW00,KSAHMSU13} we include properties of the diff (\eg, churn and diffusion measured via number of files modified), author (\eg, prior experience modifying changed file), as well as properties used in traditional file-based prediction (\eg, \cite{dalal1988should,musa,GKMS99,MW00,nagappan2005use})  such as file properties and prior \SEVs linked to the file. 

\textit{Results summary.}
Our baseline regression model captures 18.7\%, 27.9\%, and 84.6\% of \SEVs while respectively gating 
the top 5\% (weekend), 10\% (yellow), and 50\% (red) of risky diffs.

\textbf{RA 2. BERT-based model:} How well does a RoBERTa-based models capture \SEVs?
%
\StarBert is a RoBERTa-based~\cite{beller2023learning} large language model that has been pre-trained 
on millions of artifacts, such as diffs, tasks, notebooks and \SEVs at \Meta. It was designed to be 
general purpose, including classification, embedding generation, score computation (regression) and
more~\cite{karmakar2021pre,ding2022can,zhou2021assessing}.
For the purpose of gating during code freeze, labeled/annotated code diffs are used to inject task-specific
inductive bias into the pre-trained model. The goal is to train (fine-tune) the model to accurately determine
whether an unseen diff is likely to cause a \SEV.

\textit{Results summary.}
The \StarBert model only captures 0.61$\times$, 0.85$\times$, and 0.81$\times$ as many \SEVs as the 
logistic regression for the weekend, yellow, and red gating zones, respectively.

\textbf{RA 3. Generative LLMs:} The recent rise of generative models has demonstrated their ability to
supercharge various software engineering activities~\cite{fan2023large,ebert2023generative,ziegler2024measuring,yeticstiren2023evaluating,hindle2016naturalness,copilot,testpilot,codecompose,multiline,codereviewwithml}. 
Naturally, the question is can we leverage generative models for predicting diff risk scores? These models possess 
two desired properties that simpler models do not:
\begin{itemize}[leftmargin=*]
    \item[(i)] They have an innate ability to understand textual and code content. Arguably, the highest quality 
    signal on diff risk comes from the code changes in the diff, which cannot be ingested into a logistic 
    regression model. 
    \item[(ii)] Generative LLMs eliminate the need to do feature engineering. They learn these features during their
    training. On the other hand, Regression models need to be trained on features that could take significant human 
    time to define, curate and implement. 
\end{itemize}

Our research approach includes two generative LLMs:

\noindent \textbf{iCodeLlama-34B}: This model is based on the publicly available\footnote{Code Llama is a family of 
large language models for code based on Llama 2 providing state-of-the-art performance among open models, infilling 
capabilities, support for large input contexts, and zero-shot instruction following ability for programming tasks. 
Fore more information, refer to \url{https://github.com/facebookresearch/codellama}} CodeLlama-34B
model~\cite{roziere2023code}. 

\noindent \textbf{iDiffLlama-13B}: Similar to iCodeLlama-34B, this model is based on CodeLlama-13B. The main difference,
apart from its smaller size, is that in addition to code and natural language, this model has also been pre-trained on 
historical data on diffs. It can be considered ``change-aware'', as it not only knows about code but 
also \emph{code changes}.

For both models, we have two subquestions:

\textbf{3a. FM LLMs:} How well does the foundation pre-trained model capture \SEVs?

For this method, we extract embeddings from the pre-trained model and train an off-the-shelf MLP~\cite{mlp} classifier to predict the risk score for each diff. The key challenge is to extract embeddings from a generative model. As input, we provide to the LLM the diff's title, test plan and code changes.

\textit{Results summary.}
Without aligning for diff risk (aka fine tuning), the iCodeLlama-34B model only captures 0.58$\times$, 0.65$\times$, and 0.82$\times$ as many \SEVs as the logistic regression for the weekend, yellow, and red gating zones, respectively. The corresponding numbers for iDiffLlama-13B are 0.65$\times$, 0.81$\times$, and 0.90$\times$.

\textbf{3b. Risk-aligned LLMs:} Does aligning the LLM towards risk prediction allow it capture more \SEVs?

For this method, we aligned the foundation model to understand the notion of risk by fine tuning it on past diffs that have or have not caused SEVs, representing the two classes. This step teaches the model the nuances of what causes a diff to be risky. The key challenge here is to fine tune a generative model for a classification problem such as predicting diff risk.

\textit{Results summary.}
When iCodeLlama-34B is risk aligned, captures 1.26$\times$, 1.28$\times$, and 0.98$\times$ as many SEVs as the logistic regression for the weekend, yellow, and red gating zones, respectively. The corresponding number for iDiffLlama-13B are 1.40$\times$, 1.52$\times$, 1.05$\times$. 

\subsection{Contributions and Learnings}
In this paper, we make the following contributions:
\begin{itemize}
    \item[(i)] We show that generative LLMs can be utilized for predicting diff risk, inherently a classification problem. 
    We explore two possible approaches towards this: (i) embeddings from foundation model, (ii) risk alignment. In both
    approaches, we address the technical problem of aggregating embeddings from a generative model, and fine-tuning a generative model for classification, respectively.

    \item[(ii)] We show that change awareness via diff pre-training adds significant value towards model performance, as DRS is a diff-related problem. Our experiments reveal that the change-aware model iDiffLlama-13B outperforms iCodeLlama-34B despite being smaller in size, with the benefit mainly coming from its pre-training on diffs.

    \item[(iii)] We show that risk alignment further pushes the models' performance as they learn the nuances of diff risk.
    In the end, the change-aware risk-aligned model iDiffLlama-13B reaches new state-of-the-art for predicting diff risk at
    \Meta. We also show that this model generalizes effectively to predicting risk for diffs beyond its training domain.

    \item[(iv)] We discuss practical challenges associated with training and aligning generative LLMs.
\end{itemize}

The remainder of this paper is structured as follows. In the next section, Section~\ref{sec:background}, we introduce 
the idiosyncrasies of developing software at \Meta, the evolution of code freezes practices at \Meta, and how the risk 
of gated diffs is shown to the developers. In Section~\ref{sec:models} we examine the three types of risk models that 
are investigated in this paper, namely, a logic regression model, a RoBERTa-based model, and LLMs-based models. 
Section~\ref{sec:EvaluationMethod} gives details on the evaluation setup and data, and in Section~\ref{sec:results} 
we present the performance of the models in identifying \SEVs. Section~\ref{sec:t2v} discusses potential threats to 
validity of our study, and in Section~\ref{sec:discussion} we discuss our results in the context of the current 
state-of-the-art. In Section~\ref{sec:conclusions}, we present our concluding remarks.
\section{Background}
\label{sec:background}
\subsection{Software Development at \Meta}
\label{sec:swdev}
At \Meta, we develop software for both our servers and client devices, including specialized 
hardware devices. This approach allows us to have fine-grained control over versioning and
configurations, and enables us to quickly push new code updates to production. Before any code 
is deployed, it undergoes rigorous testing, including peer review, in-house user testing, 
automated tests, and canary tests. Once the code is deployed, engineers closely monitor logs to 
identify potential issues.

At \Meta, we place a strong emphasis on code reviews as part of our development process. We use
Phabricator as the cornerstone of our CI system, which facilitates modern code reviews. Developers
submit their code for review, creating a patch representing the initial version of the code.
Reviewers can suggest improvements, leading to additional revisions until the diff is either 
approved and incorporated into the codebase or rejected. This process promotes high coding 
standards, helps detect flaws, and spreads knowledge throughout the organization.

In addition to our focus on code reviews and testing, we also have a formal process for reporting 
and addressing bugs, outages, or incidents. These are reported as \SEVs (Site Events), which are 
used when the core functionality of a product is impacted without any workaround, there are numerous
and critical customer reports for the same issue, or in case of important nonfunctional problems 
related to privacy, security, pricing, or billing. By having a clear process for reporting and
addressing these issues, we can ensure that we maintain the quality and reliability of our products.
\subsection{The Evolution of Code Freeze Practice at \Meta}
Code freeze is an old practice. At \Meta, the practice of code freeze is still observed during 
certain periods of the year, such as holidays or major events. During this time, \Meta limits or 
suspends changes to its production systems to minimize the risk of service disruptions or outages.
This means that developers are not allowed to push new code to production, and any ongoing
deployments must be completed before the code freeze starts. The exact duration and scope of a 
code freeze can vary depending on the specific circumstances, but the goal is always to ensure the
stability and reliability of \Meta's systems during critical periods. While code freeze may seem 
like an old practice, it remains an important part of \Meta's development process, as it helps 
the company maintain the high availability and performance of its products and services.

Unlike traditional code freeze, where a branch is cut and only fixes are integrated, code freeze at \Meta is actually a code pause or delay, where code is not landed into the monorepo to be released for a set, and usually, short period of time. For example, over Black Friday the code is frozen for two reasons: (1) Fewer engineers were available and more disruption for SRE and on-call engineers. (2) Black Friday sees high usage of our systems and we did not want to degrade the user experience. 

The code freeze process has evolved. (1) Originally, the freeze was based on the decisions of the release engineering team. However, this does not scale~\cite{rossi2017rapid}. As a result, the decision to land a diff was pushed back to individual engineers. (2) To deal with this, \Meta decided to have experts select a set of paths and freeze diffs that touch these paths. The main drawback of relying on individual experts is that it is a perception of risk and requires constant manual effort to update. It is also not very precise including all files on a path. Furthermore, different types of change have different risk, for example, a comment only change to a file on a risky path has no risk. (3) The current solution used in production is a logistic regression model that is based on a careful selection of around a dozen  predictors from almost 100 the potential predictors handcrafted to represent 
various aspects of diff criticality, diffusion, author expertise, and modified file properties. While some of the predictors were obtained from existing literature on just-in-time defect prediction (see a recent survey~\cite{zhao2023systematic}), most were based on the specifics of \Meta's development process. The selection process and ultimate predictors are described in 
Section~\ref{s:glm}.

\subsection{Showing the Risk of Gated Diffs to the Developers}
Any diff that is above the gated risk threshold, will be blocked from landing. This means that if a diff
exceeds the risk threshold set by the team or organization, it will not be allowed to land and make 
changes to the codebase. This is done to prevent high-risk changes from being introduced into the codebase
and potentially causing \SEVs. Instead, the diff will need to be reviewed and modified to reduce its risk
score before it can be landed.

\begin{figure*}
    \centering
    \includegraphics[width=.75\textwidth]{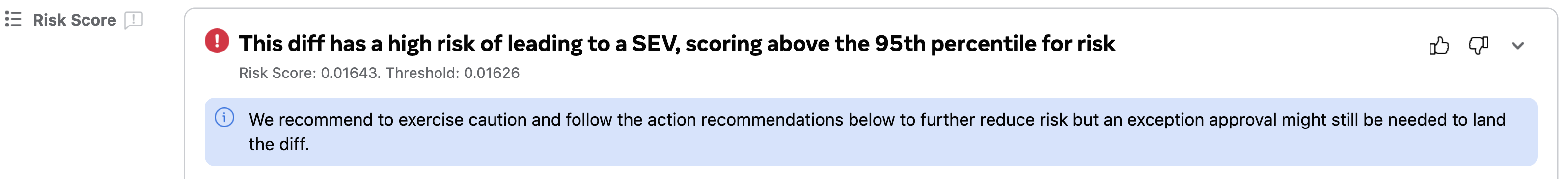}
    \caption{Showing the Risk in the Phabricator UI}
    \label{fig:phab}
\end{figure*}

The developer can wait until the code freeze is over and then land their diff. For example, if the code
freeze starts on a Friday, the developer can wait until the following Monday morning to land their diff. 
This allows the developer to ensure that any issues or problems that may arise during the code freeze are
addressed before their diff is landed. Additionally, waiting until after the code freeze is over ensures 
that the diff is not blocked from landing due to the increased risk threshold. In the event that the diff 
needs to land, there is an escalation process. There are a set of standard reasons that will escalate the 
diff to the managers that may accept the reason for the diff to land during the code freeze.

Figure~\ref{fig:phab} shows the information provided to 
the developers in the Phabricator UI, which includes the risk score of the diff, feedback mechanisms, 
informational section, reasons for the diff to be considered risky, and potential actions. The risk score 
is a value that represents the level of risk associated with the diff. The feedback mechanisms allow the
developer to provide feedback on the usefulness of the diff risk score. The reasons for the diff to be considered risky are also listed, along with potential actions that the developer can take to reduce the 
risk score and land the diff successfully. 
%
\section{Models}
\label{sec:models}
We examine three types of models: regression model, RoBERTa-based model, and LLMs.

\subsection{Regression Model}
\label{s:glm}
\label{sec:RegressionMethod}

Table~\ref{tab:defineRegression} shows the features used in one version of the the final model. 
For a logistic regression model the fitted model is simply a single vector multiplication of fitted coefficients and predictors, so the only hard part is to ensure that the values of the predictors are available in real time. 

The large number of predictors (close to 100 were calculated based on academic literature and specific aspects of the way software development at \Meta is structured -- see Section~\ref{sec:swdev}) and high correlations among them provide opportunity to select a single predictor from each of the fairly large correlated groups of predictors. Such clusters of predictors tend to be representing distinct phenomena 
that may increase the risk. For example, how diffuse the diff is and how much code it modifies, how much experience author had modifying the changed code previously, the properties of the modified files, such as programming language used, complexity of the code in the modified files, and external information such as prior history of being modified by a SEV causing diff or being used to implement highly critical service.  

This process involves considerations on prioritizing predictors that are actionable and easy to explain to developers. Obviously, any constraints on predictor selection may affect model performance, so this careful manual selection process attempts to balance actionability, ease of understanding, and performance. 

Table~\ref{tab:defineRegression} shows features of the production model presented in the results section. It includes properties of the change, like how many files are modified, properties of the modified code, like lines of code changed or programming language, properties of the developer making the change like their experience and part of organization. More unusual predictors include an indicator of whether the code is a part of a critical service. Most important software services are classified according to their criticality and code executed as part of such services is then tagged according to that class. To fit the model and do historic performance analysis 
all measures are calculated as of time the diff was landed. 

\begin{table*}
\caption{The features used in the logistic regression that is in production}
\label{tab:defineRegression}
\centering
\begin{tabular}{ll}









\hline
Feature type	&	Feature used in Logistic Regression	\\ \hline

Diff 		
& $log$ of the added and deleted SLOC relative to size of file (ratio)\\
& New files created by the diff (boolean) \\
& Diff only creates new files (boolean) \\
\hline

Diffusion & $\log$ of the number of files in this diff	\\ 
& $log$ of the number of authors that modified changed files	\\ \hline
\hline

Criticality & Previous SEV in the file (boolean) \\ 
& Previous SEV in the folder (boolean) \\
& Is file involved in high-criticality service (boolean) \\
\hline

File & Total logical complexity of files touched in this diff \\ 
 & Programming language (seven boolean indicators if at least one file in that language is modified) \\ 
\hline

Expertise	& If the author is the original creator of the file \\
        & Number of diffs previously landed by the author\\
\hline

\end{tabular}
\end{table*}

\begin{table*}
\caption{The features fed to the LLMs during both training and inference for DRS}
\label{tab:llmFeatures}
\centering
\begin{tabular}{ll}

\hline
Feature type	&	Feature fed to the LLM	\\ \hline

Diff Title		
& Title of the diff, typically a concise description of the code change in a few words\\

Test Plan
& Commands (build, lint, tests) executed by the diff author to validate the code changes \\

Code changes
& Filenames and the corresponding code changes in the standard unified diff (``unidiff'') format \\
\hline

\end{tabular}
\end{table*}
\subsection{Why Large Language Models?}
The underlying DRS model today is a logistic regression trained on diff metadata such as the diff, file, author and related features. One limitation of such simple models is that they do not understand any content based features, e.g., diff code, summary test plan etc., which arguably can contain the highest signal of diff risk. For instance, a diff that only adds comments to a file can never cause a \SEV, but could be flagged by the model because it touches the same file as a previous \SEV-causing diff. On the other hand, a particular code pattern that consistently causes a \SEV could be missed by the model because of its inability to understand code changes.

Recently, LLMs have been proven to be effective in understanding and generating textual and code content. As a result they have been widely adopted to help improve software engineering productivity~\cite{nguyen2022empirical,codereviewwithml,schafer2023empirical,schafer2023adaptive}

Large language models (or LLMs) in recent years have been proven to be efficient in code understanding, improvement and generation, thus starting to be widely adopted to help improve software engineering efficiency[1,2]. We would like to leverage LLMs to understand the code change when doing diff risk score analysis to improve the accuracy of diff risk score, boosting engineering productivity by highlighting and providing suggestions for engineers.

LLMs also eliminate the potentially expensive feature engineering that is needed by simpler models such as logistic regression. For instance, when dealing with code, LLMs do not need to be fed hand-curated features about what language the code is in, how large it is, whether it is a comment, and so on. They can automatically learn these features internally as part of their training, albeit not human interpretable. Table~\ref{tab:llmFeatures} lists the textual and code features from a diff that are fed to the LLMs for both training and inference for risk prediction.
\subsection{\StarBert  model}
\label{sec:StarBertMethod}
\StarBert is a RoBERTa-based model for understanding semantics of various artifacts across \Meta’s internal platform~\cite{beller2023learning}. It can be used for several different purposes, including classification, embedding generation, score computation (regression) and more~\cite{karmakar2021pre,ding2022can,zhou2021assessing}.

At the core of StarBERT platform are the pretrained models. 
The representations learned by these models are generic and transfer to downstream tasks of platform users through fine-tuning. Transfer learning provides substantial benefits to user teams by (1) requiring much smaller labeled datasets, (2) immediate benefit of base-line performance out of the box, (3) simplifying feature selection and (4) considerably faster model iterations. Pretraining is essential to save on the research and development that goes into model building and selection as well as enable impact that is otherwise not possible.

To fine tune the model for a specific task such as computing the risk score of a diff, labeled/annotated code diffs are used to inject task-specific inductive bias into the pre-trained model. The goal is to train (finetune) the model to accurately determine whether an unseen diff is likely to cause a \SEV.
\subsection{Generative LLMs}
\label{subsec:LLMmodels}
The state-of-the-art generative LLMs of today have been effective at powering several software engineering tasks, 
ranging from code completion~\cite{copilot,codecompose}, test generation~\cite{testpilot}, to code
review~\cite{codereviewwithml}. These models are based on a ``foundation model'' that is pre-trained on internet-scale
data (billions of tokens) on a simple generic task, \ie, next token prediction. They are then prompted as-is or 
fine-tuned further on domain specific data to elicit increased performance on a specific task.

CodeLlama~\cite{roziere2023code} is such an AI model built on top of Meta GenAI's Llama 2~\cite{touvron2023llama}, 
fine-tuned for generating  and discussing code. 
Essentially, CodeLlama features enhanced coding capabilities. 
It can generate code and natural language about code, from both code and natural language prompts
(e.g., ``Write me a function that outputs the Fibonacci sequence''). It can also be used for code completion and 
debugging. It supports many of the most popular programming languages used today.

For this work, we considered two models built on top of CodeLlama.
\subsubsection{iCodeLlama}
\label{sec:iCodeLlamaMethod}
iCodeLlama is a \Meta-internal model that we have developed by further pre-training CodeLlama on \Meta's 
proprietary code base, including code from our monorepo, natural language comments appearing inline, and 
other code-related documentation. It exhibits similar capabilities as CodeLlama but on \Meta's internal 
contexts. iCodeLlama-7B has been used to power \Meta's code completion system, CodeCompose~\cite{codecompose,multiline}.

\subsubsection{iDiffLlama}
\label{sec:iDiffLlamaMethod}
iDiffLlama is also based on CodeLlama, but in addition to pre-training on code and natural language, it is also 
pre-trained on internal diffs, \ie, code changes. This makes the model ``change-aware'' as it understands the 
nuances of not just static code, but also the evolution of code through changes. Particularly, diffs represent 
an intent by a developer to commit their changes to the monorepo. iDiffLlama captures that intent through 
learning the relationship between a diff's title, summary, test plan and the corresponding code changes.
\subsection{Using Generative LLMs for classification}
The first question to answer when using today’s LLMs (iCodeLlama or iDiffLlama) for a problem like DRS is, 
can generative models be used for classification problems? They are designed for generating text 
or code and having conversations rather than classifying an input into categories. 

One simple way to 
leverage a GenAI model for classification is through prompting (0-shot or few-shot). While this approach 
works for other applications, there are two problems with it particular to DRS:

\begin{itemize}
\item The model needs to provide a risk score rather than a binary label, as the score is used for ranking. Prompting the model to output a score is quite tedious as the model has to generate a numerical score token by token. It is also unreliable, as the model does not have a universal view of risk and may generate uncalibrated scores across different examples.
\item The input to DRS is not a simple piece of text but a heavyweight diff which often occupies a significant portion of the context window, making few-shot with current generative models infeasible. DRS is also an inherently hard problem due to its needle-in-the-haystack nature, and it is difficult to provide sufficiently balanced examples for few-shot learning.
\end{itemize}

Instead, we explored two methods for using LLMs for classification: embeddings from foundation model, and risk alignment.
\subsubsection{Embeddings from foundation model}
\label{sec:EmbeddingsMethod}
\begin{figure*}
    \centering
    \includegraphics[width=.75\textwidth]{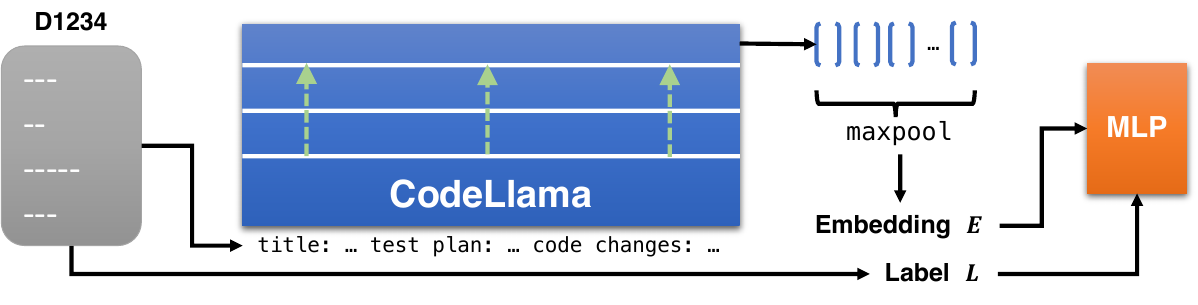}
    \caption{Pipeline: Embeddings from foundation model. A model forward pass is run on the input diff, and hidden states from the final Transformer layer are aggregated via maxpool to form the diff's embedding $E$. An external classifier is then trained on the embeddings and labeled DRS data.}
    \label{fig:embeddings}
\end{figure*}

The idea here is to extract the relevant features from a given diff, embed the features using the pre-trained foundation LLM, and use those embeddings along with the labeled DRS data to train an off-the-shelf classifier such as an MLP. The pipeline is illustrated in Figure~\ref{fig:embeddings}.

To compute embeddings for an input diff, we do a single forward pass on the input and aggregate the hidden states in the final layer of the Transformer model. We experimented with meanpool and maxpool, and found maxpool to work slightly better. One advantage of this approach is that it makes it easier to integrate the embeddings with the current logistic model trained on other metadata features.

\subsubsection{Risk alignment}
\label{sec:RiskAlignMethod}
\begin{figure*}
    \centering
    \includegraphics[width=.75\textwidth]{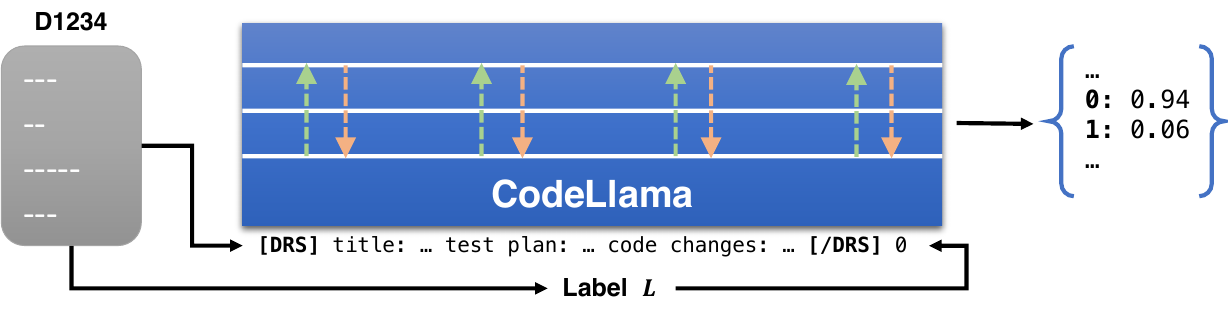}
    \caption{Pipeline: Risk alignment. The LLM undergoes supervised fine-tuning (SFT) on labeled DRS data, where each diff is annotated with special tokens \texttt{[DRS][/DRS]} that denote the model to predict risk. The LLM is trained to generate the label token 0 or 1 appended to the end of the input. During inference, the risk score is computed based on the token probabilities of the labels 0 and 1.}
    \label{fig:sft}
\end{figure*}

In this approach (depicted in Figure~\ref{fig:sft}), we run a typical supervised fine-tuning (SFT) phase on the pre-trained LLM using labeled DRS data. In order to transform the classification problem into a generative task, we use special markers \texttt{[DRS][/DRS]} to annotate the input, append the label (0 or 1) to the end, and train the model to ``generate'' the label token.

During inference, we can query the model to generate the label token. To get the risk score rather than just the label, we extract the probability of the label tokens ``0'' and ``1'' from the next token distribution. If the model is aligned well, all other tokens in the model's vocabulary should have close to zero probability of being generated.

The advantage of this approach is that it allows the LLM to backpropagate and learn the nuances of diff risk, \ie, makes it ``risk-aligned'', rather than serving untuned embeddings that only capture general properties of diffs. However, this requires a computationally expensive fine-tuning of the LLM and deployment of this specialized model.

\section{Evaluation Method and Data}
\label{sec:EvaluationMethod}
To ensure a fair comparison of different models for Diff Risk Scoring (DRS), we use the same dataset and splits
across all models. We split our data chronologically (by diff closed date) into training, validation, and testing
sets. This approach offers a more realistic split compared to random sampling, as it better reflects the temporal
nature of the data.

Although \Meta has a monorepo, we scoped our experiment and data to a particular organization's diffs within the company. This allows us to test the model's generalization capability (see Section~\ref{subsubsec:gen}), as well as focus engineering efforts on deployment within the pilot organization at \Meta.

Our testing data consists of $181052$ regular diffs and $305$ \SEVs. By using a chronological split, we can evaluate
the models' performance on unseen data that was not used during training or validation. This allows us to better
assess their ability to generalize to new, unseen data in real-world scenarios.

\begin{table}
\caption{Data splits}
\label{tab:defineRegression}
\centering
\resizebox{\columnwidth}{!}{%
\begin{tabular}{lllll}
\hline
             & diff closing data	from &	diff closing data to & sample size & \SEV count/rate	\\ \hline
Training     & 2022-01-01 & 2023-05-04 & 855282 & 1981 (0.23\%) \\
Validation   & 2023-05-05 & 2023-05-06 & 120967 & 214  (0.18\%) \\
Testing      & 2023-07-01 & 2023-10-02 & 181052 & 305  (0.17\%) \\
\hline
\end{tabular}
}
\end{table}

\text{Dealing with skewed data.} When either training an external classifier or aligning the LLM for risk prediction, there is an extreme imbalance in the labeled data, as SEV-causing diffs account for less than 1\% of all diffs (see Table~\ref{tab:defineRegression}). To deal with this imbalance we resampled the examples in the training data to arrive at a 5:1 negative-positive class ratio. We did this only for the training data to ensure that the model does not regress to overwhelmingly predicting the majority class. We did not modify the validation and testing data in order to preserve the real world distribution. We then froze the resampled training data and used it for all models.


For the dataset we select the top g riskiest diffs from each model, and determine as outcome what percentage of \SEVs that would be captured. We use the current production gating levels of 5\% (weekend), 10\% yellow or intermediate, and 50\% red or high.

\section{Results}
\label{sec:results}

\begin{table*}
\centering
\caption{The percentage of captured SEVs by model as well as the percentage increase relative to the current project logistic regression model}
  \begin{tabular}{lrr|rr|rr}
\toprule

Model	&	\multicolumn{2}{c}{Weekend (g = 5\%)}			&	\multicolumn{2}{c}{Yellow (g = 10\%)}			&	\multicolumn{2}{c}{Red (g = 50\%)}				\\ 
	&	SEVs Captured	&	vs Regression	&	\% SEVs Captured	&	vs Regression	&	\% SEVs Captured	&	vs Regression		\\ \hline
Logistic Regression 	&	18.7	\% &	$-$	$\times$ &	27.9	\% &	$-$	$\times$ &	84.6	\% &	$-$	$\times$ 	\\ \hline
StarBERT	&	11.5	\% &	0.61	$\times$ &	23.6	\% &	0.85	$\times$ &	68.9	\% &	0.81	$\times$ 	\\ \hline
iCodeLlama-34B  	&	10.8	\% &	0.58	$\times$ &	18.0	\% &	0.65	$\times$ &	69.2	\% &	0.82	$\times$ 	\\
iCodeLlama-34B risk aligned	&	23.6	\% &	1.26	$\times$ &	35.7	\% &	1.28	$\times$ &	83.0	\% &	0.98	$\times$ 	\\ \hline
iDiffLlama-13B  	&	12.1	\% &	0.65	$\times$ &	22.6	\% &	0.81	$\times$ &	75.7	\% &	0.90	$\times$ 	\\ 
\textbf{iDiffLlama-13B risk aligned}	&	\textbf{26.2}	\% &	\textbf{1.40}	$\times$ &	\textbf{42.3}	\% &	\textbf{1.52}	$\times$ &	\textbf{88.5}	\% &	\textbf{1.05}	$\times$ 	\\

    \bottomrule
  \end{tabular}
  \label{tab:SEVsCapturedResults}
\end{table*}

\subsection*{RA 1. Logistic Regression}
\textit{How well does the current model capture SEVs?}

For production use at scale, we need a simple model and features that can be calculated across thousands of engineers and diffs. 
As we discussed in the model development, see Section~\ref{sec:RegressionMethod}, we reviewed the literature on defect modelling and selected the represented set of features shown in Table~\ref{tab:defineRegression}. 

In Table~\ref{tab:SEVsCapturedResults}, we see that our baseline regression model captures 18.7\%, 27.9\%, and 84.6\% of SEVs while respectively gating the top 5\% (weekend), 10\% (yellow), and 50\% (red) of risky diffs.
This regression model has been used in production for more than 9 months.

\subsection*{RA 2. \StarBert}
\textit{How well does a RoBERTa-based model capture SEVs?}

\StarBert is a RoBERTa-based~\cite{beller2023learning} large language model. 
For the purpose of gating during code freeze, labeled/annotated code diffs are used to inject task-specific inductive bias into the pre-trained model. 
The goal is to train (finetune) the model to accurately determine whether an unseen diff is likely to cause a SEV. The model develop in fully described in Section~\ref{sec:StarBertMethod}.

In Table~\ref{tab:SEVsCapturedResults}, we see the \StarBert{} model only captures 0.61$\times$, 0.85$\times$, and 0.81$\times$ as many SEVs as the logistic regression for the weekend, yellow, and red gating zones, respectively.
Clearly the RoBERTa-based model cannot replace the logistic regression for production purposes. 

\subsection*{RA 3. Generative LLMs}

For generative LLMs, we pick the two models. Specifically, we take iCodeLlama-34B and iDiffLlama-13B, having 34B and 13B parameters, respectively. 
The full details on the model development can be found in Section~\ref{subsec:LLMmodels}. 

\subsubsection*{RA 3a. FM LLMs} 
\textit{How well does the foundation pre-trained model capture SEVs?}

We use the embedding in the from the iCodeLlama-34B and iDiffLlama-13B and use a classifier to determine the DRS risk score. The method is described in Section~\ref{sec:EmbeddingsMethod} and the pipeline is shown in Figure~\ref{fig:embeddings}.
We trained an external 3-layer MLP classifier on the embeddings consisting of $(100,150,50)$ hidden units, respectively. 

In Table~\ref{tab:SEVsCapturedResults}, we see that without aligning for risk, the iCodeLlama-34B model only captures 0.58$\times$, 0.65$\times$, and 0.82$\times$ as many SEVs as the logistic regression for the weekend, yellow, and red gating zones, respectively. The corresponding numbers for iDiffLlama-13B are 0.65$\times$, 0.81$\times$, and 0.90$\times$. 
This shows that unaligned foundation LLMs are not effective at capturing SEVs.

\subsubsection*{RA 3b. Risk aligned LLMs} 
\textit{Does aligning the LLM towards risk prediction allow it capture more SEVs?}

We saw that using the embeddings with an MLP was ineffective. We fine-tune the iCodeLlama-34B and iDiffLlama-13B by running one epoch on the training data with an effective batch size of 64. The training was conducted on a cluster of 64 Nvidia A100 GPUs, and took around 2-4 hours to complete depending on the size of the LLM. The full methodology to make the LLMs risk aligned is described in Section~\ref{sec:RiskAlignMethod} and the pipeline is shown in Figure~\ref{fig:sft}.

In Table~\ref{tab:SEVsCapturedResults}, when iCodeLlama-34B is risk aligned it captures 1.26$\times$, 1.28$\times$, and 0.98$\times$ as many SEVs as the logistic regression for the weekend, yellow, and red gating zones, respectively. The corresponding number for iDiffLlama-13B are 1.40$\times$, 1.52$\times$, 1.05$\times$. 

An interesting observation here is that while both risk aligned models outperform the baseline logistic regression, iDiffLlama-13B outperforms iCodeLlama-34B despite being a smaller model. This reveals that change-aware (diff) pre-training adds significant value to an LLM's performance on predicting diff risk. This makes sense as DRS is essentially a diff related problem. Thus, we have arrived at a new state-of-the-art for diff risk prediction at \Meta, with the change-aware risk-aligned model iDiffLlama-13B. 

\subsubsection*{3c. Generalizability} 
\label{subsubsec:gen}
\textit{How well does the best performing research approach generalize beyond its organization?}


In order to test the model's generalizability, we used the risk aligned iDiffLlama-13B to predict the diff risk for diffs in another organization under \Meta. We ensured that there is no overlap between diffs that land in the two organizations' part of the code base. As before, we compared the model to an existing logistic regression model used in the other organization.

Out of 160 SEVs the logistic regression model was able to capture 11.9\%, 26.3\% and 84.4\% of SEVs at the three gating thresholds of 5\%, 10\% and 50\%, respectively. However, iDiffLlama-13B was able to capture 21.9\%, 39.4\%, and 88.8\% of SEVs at the same thresholds, translating to a 1.84$\times$, 1.5$\times$, and 1.05$\times$ improvement over the baseline. This shows that LLMs are effective at generalizing beyond their training domain for predicting diff risk.

\section{Threats to Validity}
\label{sec:t2v}

\subsection{Generalizability}
One potential threat to the generalizability of our findings is the limited scope of our dataset. Our study 
was conducted using data from a single company, which may not be representative of all software development
scenarios. The specific tools, processes, and culture of our company may have influenced the results of our 
study, and other companies may have different factors that impact their release engineering decisions. 
Therefore, it is unclear whether our findings can be generalized to other contexts.
\subsection{Construct Validity}
Another potential threat to validity is construct validity. Our study focused on diff risk scoring for \SEV
prevention only, and did not consider other factors that may impact release engineering decisions, such as 
business value or user feedback. Therefore, our approach may not fully capture the complexity of the problem 
and may not be suitable for all release engineering scenarios. Additionally, our models were trained on 
historical data, which may not be representative of future software development scenarios. Therefore, the 
performance of our models may degrade over time and require periodic retraining.
\subsection{Internal Validity}
Finally, there are potential threats to internal validity in our study. One potential threat is the evaluation 
metric used to assess the performance of our models. We evaluated our models based on a single metric, \SEV 
capture rate, which may not fully capture the complexity of the problem. 
Other metrics, such as 
the ultimate goal of reducing the number and impact of \SEVs and minimizing disruptions to developer productivity 
from gating, may provide additional insights into the effectiveness of our approach. Another
potential threat is the possibility of unmeasured confounding variables that may have impacted our results. For
example, there may be other factors that influence the likelihood of a diff causing a \SEV, such as code quality 
or team dynamics, that were not captured in our analysis.

%
\section{Discussion and Literature}
\label{sec:discussion}
\subsection{Release Engineering}
Release engineering is a discipline of software engineering that focus on creating pipelines that takes source code and turns it 
into a final product that is ready for release. This includes compiling, packaging, testing, and signing the 
code~\cite{adams2015practice}. The goal of release engineering is to ensure that the pipeline is efficient and reliable, so that
products can be delivered to customers quickly and with high quality. This is especially important for companies like \Meta, 
providing services to  millions of users, as they need to be able to deliver updates to their customers in a timely manner. 
Code freeze, the main focus of this paper, is a common practice during which no changes are allowed to be made to the codebase, 
in order to ensure stability and prevent any issues during critical periods such as holidays or major events.

The models we discussed in the paper aim at relaxing the code freeze periods, this way allowing users to land diffs that are 
considered not to harm the stability of the current release. This is, a move towards a code chill, which is a step in moving away 
from a hard/strict code freeze, as it allows teams to deploy important changes more freely and gives them the autonomy -- and 
responsibility -- to make the call on whether a deployment is important enough to risk deploying. To help teams have that 
autonomy, and responsibility, we are equipping the teams at \Meta with a quantification of the diff risk score of their diffs. 

The models presented in the paper aim to reduce the strictness of code freeze periods, enabling developers to land diffs that 
are deemed safe for the current release (that is, do not compromise the stability of the current release). This approach
represents a shift towards a ``code chill,'' which is a step towards relaxing strict code freeze policies and allows teams 
to deploy changes more easily while assuming responsibility for their decisions. To support this autonomy and accountability, 
we are providing \Meta teams with a quantitative assessment of the risk associated with their diffs.
\subsection{Defect Prediction Models}
Code quality is one of the pillars of software engineering and statistical models attempting to predict software defects have 
been developed in the previous century~\cite{musa,GKMS99}. While traditional defect prediction models focus on identifying 
which files will have defects, just-in-time defect prediction models~\cite{MW00,KSAHMSU13} attempt to predict which change 
(diff) will cause a problem. Both fields are extremely mature (see systematic surveys of recent work in, 
\eg, ~\cite{zhao2023systematic} for just-in-time prediction, ~\cite{zain2023application} for the more recent development of 
application of  deep learning methods).

One aspect that distinguishes our effort from almost all prior work is the accuracy of identifying \SEVs causing diffs. 
At \Meta, all \SEVs undergo a rigorous review process and the trigger (cause) is identified. While sometimes \SEV is 
caused by hardware failures or network outages, often the cause is a software change (diff). Virtually all published work 
uses rather noisy estimates of which diff caused the problem based on information from issue-fixing diffs, while in our 
case the data is manually derived by area experts. 
%
\subsection{Costs: Human vs. Machine}
There is always a cost associated with training and deploying models. For the logistic regression model, the cost is in terms 
of human time involved in feature engineering, \ie, cleaning the data, curating and implementing features, determining best 
predictors, feature importance, etc. Whereas, the compute cost involved in training and deploying the model is relatively small. 
They can be trained on CPUs, and usually take in the order of minutes to complete.

On the other hand, LLMs require heavy compute power to operate. They require GPUs to train and deploy, and take in the order 
of hours or days to train. However, they completely eliminate the need to do any feature engineering. They are able to take 
in raw data and learn these (latent) features during their training. They also have a highly desirable property of transfer 
learning across training phases, as they are based on a powerful foundation model pre-trained on billions of tokens. These 
characteristics allow LLM-based solutions to follow simpler scaling laws, as data and compute become the only two requirements 
for scaling up.
%

%
\section{Future Work}
We discuss two avenues of future work: (1) generative reasoning on why a diff may lead to a SEV (2) ensemble learning combining LLMs and regressions. 
\subsection{LLM fine-tuning for generative reasoning}
Generative reasoning is an alternative approach to classification. With an appropriate SFT phase, LLMs could potentially synthesize an answer of whether a diff is risky along with an explanation of why, in a purely generative manner. This method is used in the AI community to have LLMs adapt to specific problem domains, such as the work done in Microsoft Research~\cite{msrGenerativeReasoning}. On the other hand, SFT for generative reasoning than classification is a much longer term effort, owing to the following:

\begin{itemize}
    \item It requires large investment to collect actual SEV reasoning traces and their root causes, rather than the label of whether the diff caused a SEV. Moreover, discussions around SEVs are multi-modal, involving a combination of natural language, code, screenshots or videos, and so on. Developing a multi-modal LLM capable of reasoning about SEVs is beyond the scope of this paper.
    
    \item Hallucinations are a major problem with generative reasoning, which would confuse the user even more if the LLM generated the wrong explanation. Mitigating hallucinations is still very early research, and the industry has little experience on this. There are existing efforts to optimize prompt engineering to mitigate hallucinations, which from our observation provides limited help.

\end{itemize}
\subsection{Ensembling LLM's and Logistic Regression's risk score}
%
%
The diff, author, and file related features in the logistic regression model continue to be powerful and valuable predictors. Consequently, it is advantageous to ensemble the LLM and the logistic regression model. We incorporated the LLM score as an additional feature into the logistic regression model. During the inference process, an LLM score is initially obtained, followed by the computation of logistic regression. In instances where the LLM score is absent (due to timeout, etc.), we impute it as the mean. 

Alternative ensemble strategies include:
\begin{itemize}
    \item Utilizing the LLM embedding as input variables for logistic regression rather than the prediction score, which may encompass more comprehensive information.

    \item  Implementing a weighted average of the LLM score and logistic regression score, where the weight is determined by maximizing precision/recall on the evaluation dataset.

    \item Considering the addition of features in logistic regression into the LLM, enabling the LLM to fully supplant the logistic regression model.
\end{itemize}

\section{Conclusion}
\label{sec:conclusions}
In conclusion, our study has demonstrated that the use of machine learning models can significantly improve the 
accuracy of diff risk scoring, which can help release engineers make more informed decisions about which diffs 
to gate. The logistic regression outperformed the RoBERTa-based models. However, the generative LLM models showed promising results, with the iDiffLlama-13B model capturing the most \SEVs among all models tested. 

%
The following are some highlights and learning worthy of note from our experiment:
\begin{itemize}
    \item Our experiment answered the question of whether generative models can be suited for a classification problem like DRS. We validated using both the foundation models and risk aligned models for classification. Although expected, the results confirm that fine-tuning a model for a problem domain significantly boosts its performance compared to using a pre-trained model, \ie, risk aligned models outperformed pre-trained models across all metrics.
    
    \item Directly using the LLM for prediction performs better than using an external classifier trained on foundation LLM embeddings. This is likely due to the fact that there is loss of information when aggregating (pooling) hidden states for embeddings, and there is no self-attention in the external classifier when making the prediction.
    

    \item The change-aware model iDiffLlama-13B outperforms the general purpose code LLM iCodeLlama-34B, despite the former being a smaller model. This is due to it being a specialist model specifically trained to understand diff data in the form of patches.
\end{itemize}

These findings highlight the potential of machine learning models to enhance the efficiency and safety of the software development process. Future work includes further refinement of the models and exploration of additional features to incorporate into the models.
%
%
%
\balance
\bibliographystyle{IEEEtran}
\bibliography{all}

\end{document}